\documentclass[amsmath,amssymb,amsbsy,prb,twocolumn,groupedaddress,showpacs]{revtex4-1}

\usepackage{bm}
\usepackage{xcolor}
\usepackage[utf8]{inputenc}
\usepackage{graphicx}
\usepackage{hyperref}
\usepackage{subfigure}
\usepackage{wasysym}
\usepackage{tabu}
\usepackage{amsmath,tabu}
\usepackage{amsfonts}
\usepackage{braket}
\usepackage{dsfont}

\newcommand{\bk}{\mathbf{k}}

\newcommand{\br}{\mathbf{r}}
\newcommand{\bd}{\mathbf{d}}
\newcommand{\re}{\mathrm{e}}
\newcommand{\dg}{\dagger}

\hypersetup{colorlinks=true,linkcolor=blue,urlcolor=blue,citecolor=blue}

\begin{document}

\title{Supersymmetry in an interacting Majorana model on the kagome lattice}
\author{Chengshu Li}
\email{chengshu@phas.ubc.ca}
\author{\'{E}tienne Lantagne-Hurtubise}
\email{lantagne@phas.ubc.ca}
\author{Marcel Franz}
\affiliation{Department of Physics and Astronomy \& Stewart Blusson Quantum Matter Institute, University of British Columbia, Vancouver, British Columbia, Canada V6T 1Z1}

\begin{abstract}
We construct a supersymmetric model of interacting Majorana fermions on the kagome lattice. In the infinite-coupling limit, the model exhibits an extensively degenerate ground state manifold separated in two topological sectors, in addition to two parity supersectors. An exact solution for thin-torus geometries allows us to analytically construct the entire zero-energy ground state manifold. Upon inclusion of hopping terms with energy scale $g$, the supersymmetry is spontaneously broken with the ground state energy density scaling as $g^4$. We also briefly discuss the non-interacting limit of the model, which exhibits a zero-energy flat band and Majorana Chern bands.
\end{abstract}

\date{\today}

\maketitle

\section{Introduction}
Supersymmetry (SUSY), a symmetry which relates bosons and fermions~[\onlinecite{weinberg_2000}], was originally proposed in high-energy physics as a possible extension of the Standard Model of particle physics~[\onlinecite{Volkov1973, Salam1974, Wess1974, Witten1981, Witten1982}]. Whereas its status as a fundamental symmetry of nature is debated, realizing SUSY as a low-energy property in condensed matter systems has attracted considerable recent interest.

Two main avenues of research have been explored so far. The first idea focuses on supersymmetric quantum critical points~[\onlinecite{Lee2007}] occurring, for example, at surfaces (edges) of 3D (2D) topological insulators~[\onlinecite{ponte_2014, Grover2014, Witczak-Krempa_2016, Jian_2017, Li2017, Li2018b}], in Weyl or Dirac semimetals~[\onlinecite{Jian2015}], and in 1D chains of interacting Majorana fermions~[\onlinecite{Rahmani2015b}, \onlinecite{2018Ebisu}] where the tri-critical Ising point~[\onlinecite{Friedan1984}] is realized. These proposals realize an emergent version of SUSY, which is not present in the underlying microscopic theory.

The other avenue is to construct a quantum-mechanical model with explicit SUSY. This corresponds to spacetime SUSY in (0+1)D -- that is, only in the time direction. In that case, SUSY is not an emergent property appearing at a critical point, but rather embedded directly in the Hamiltonian describing the system, which is constructed by squaring a ``supercharge operator". This strategy has a long history in various lattice models~[\onlinecite{Nicolai1976,Fendley2003, Fendley2005, Yu2008, Huijse2008, Yu2010, Huijse2012, Bauer2013, Sannomiya2017_PRD}]. Recently, interest in this approach was revived by considering systems comprising Majorana fermions instead of complex fermions. The main advantage of using Majorana operators to construct a supercharge is that they square to identity~[\onlinecite{Elliott2015}], often resulting in simpler Hamiltonians. This led to recent proposals for 1D Majorana chains which admit a supersymmetric phase with an exactly solvable point~[\onlinecite{Fendley2018}, \onlinecite{Sannomiya2019}].

\begin{figure}
    \centering
    \includegraphics[scale=0.5]{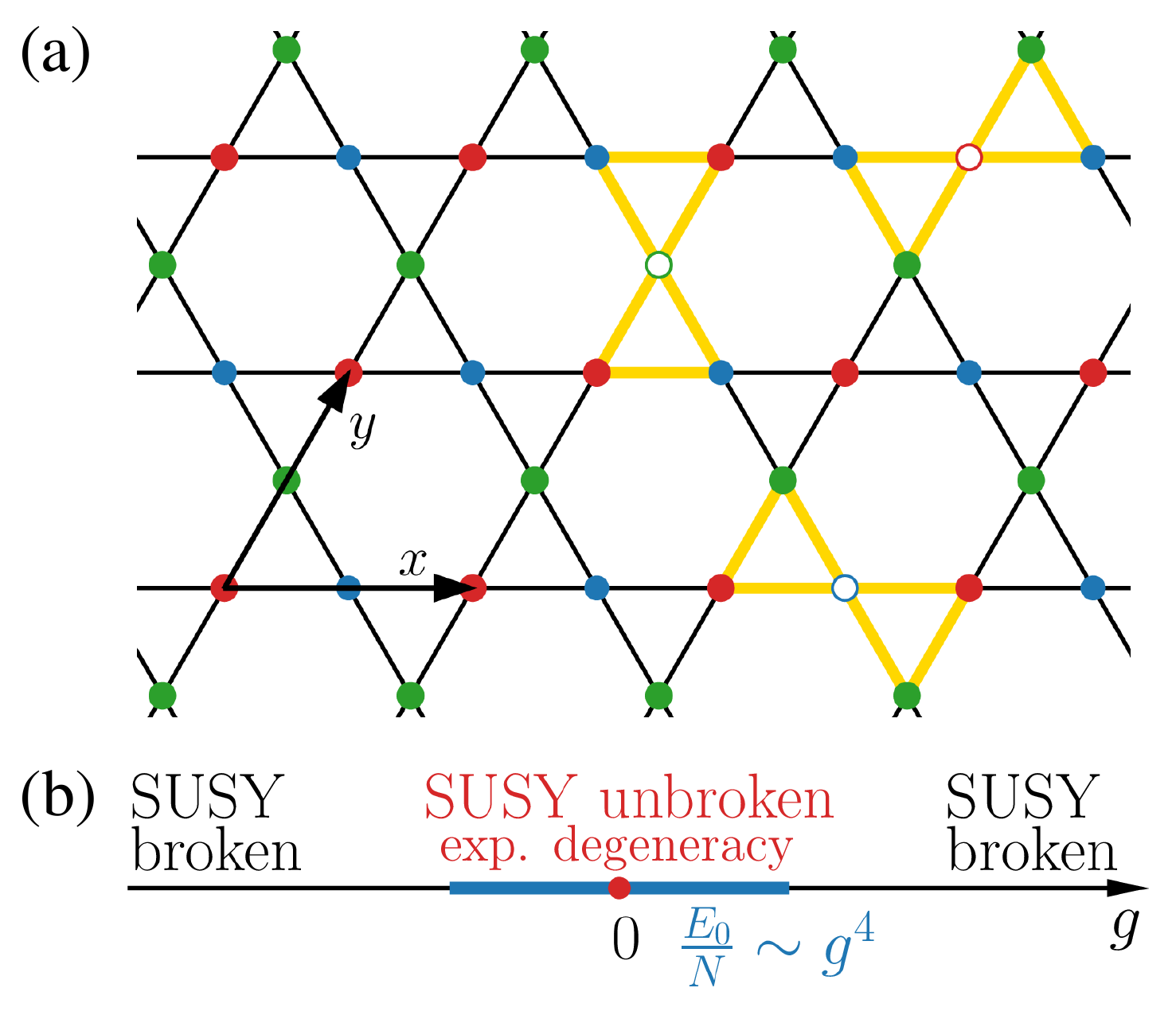}
    \caption{(a) The kagome lattice, where the three sublattices are colored red (r), green (g) and blue (b). The supersymmetric Hamiltonian [Eq.~(\ref{eq:Hamiltonian})] comprises nearest-neighbor bilinear terms (with strength $g$) and four-fermion bowtie interactions, sketched in gold. (b) The phase diagram of the model obtained in the thin-torus limit.}
    \label{fig1}
\end{figure}

Here, we present an extension of these ideas to 2D systems of interacting Majorana fermions~[\onlinecite{Chiu2015}, \onlinecite{Rahmani2018}], which were recently studied (in a different context) on the square~[\onlinecite{Affleck2017,Kamiya2017,Wamer2018}] and honeycomb~[\onlinecite{vijay2015}, \onlinecite{Li2018}] lattices. We construct a supersymmetric model of interacting Majorana fermions on the kagome lattice, Fig.\,\ref{fig1}(a), with strictly local two-fermion and four-fermion terms and a single parameter $g$ controlling their relative strength. Its phase diagram is shown in Fig.\,\ref{fig1}(b). In the infinite-coupling ($g=0$) and thin-torus (quasi-1D) limit, the model exhibits unbroken SUSY and a ground state manifold with degeneracy that is exponential in system size, a property sometimes referred to as \emph{superfrustration}~[\onlinecite{Fendley2005}].  
Furthermore, this limit admits an \emph{exact solution} which allows us to explicitly construct all ground states and uncover the presence of two topological sectors. Away from the infinite-coupling limit, numerical calculations indicate that SUSY is (weakly) spontaneously broken, with a ground state energy density scaling as $E_0/N \sim g^4$.

The rest of this paper is organized as follows. In Sec.~\ref{sec:model} we introduce the model and discuss its symmetries. In Sec.~\ref{sec:non-interacting} we solve the non-interacting ($g = \infty$) limit of the model. In Sec.~\ref{sec:inf-interacting} we discuss the infinite-coupling limit ($g=0$) and its exact solution for the ground state manifold in the thin-torus limit. Finally, in Sec.~\ref{sec:finiteg} we analyze SUSY breaking for $g \neq 0$ using a combination of numerical techniques: exact diagonalization, infinite density-matrix renormalization group (iDMRG), and degenerate perturbation theory.

\section{The model}
\label{sec:model}

We consider the kagome lattice with one Majorana fermion at each site [Fig.\,\ref{fig1}(a)]. We then define the supercharge as the following Hermitian operator,
\begin{align}
    Q = g \sum_j \chi_j + \sum_{\Delta} V_\Delta, \label{eq:1}
\end{align}
where $V_\Delta = i \prod_{j \in \Delta} \chi_j $ are plaquette operators on each triangle, and the product over $j$ is taken clockwise. The supercharge operator can also be defined with a relative minus sign for upwards-facing and downwards-facing triangles, leading to similar physics (see Appendix~\ref{app:general} for details). The usual algebra for Majorana fermions is $\{ \chi_i, \chi_j \} = 2 \delta_{ij}$ and $\chi_j^\dagger = \chi_j$. We construct the Hamiltonian by squaring the supercharge operator, $H = Q^2$, leading to
\begin{align}
    H &= 2 i g \sum_{\langle i,j\rangle} \chi_i \chi_j + 2 \sum_{\bowtie} \left( \prod_{j \in \bowtie} \chi_j \right) + \frac{2N}{3}  +  N g^2,
      \label{eq:Hamiltonian}
\end{align}
where $N$ is the number of lattice sites. The first term describes nearest-neighbor hoppings, where the pairs are ordered in the clockwise direction along each triangle. The second term represents four-fermion interactions involving one Majorana fermion on each corner of the bowtie structures of the kagome lattice [see Fig.\,\ref{fig1}(a)]. Note that the Majorana fermion on the central site of each bowtie (shared by two triangles) is absent, a consequence of the property $\chi_j^2 =1$. The Hamiltonian is positive-semidefinite by construction -- $H$ and $Q$ being mutually commuting Hermitian operators, we can choose common eigenstates with real eigenvalues $E$ and $q$, respectively, which gives $E = q^2\geq0$. 

This procedure, whereby one constructs a single Hermitian supercharge and then squares it, always generates a Hamiltonian with explicit $\mathcal{N} = 1$ SUSY. However, as with regular symmetries, SUSY can be broken spontaneously. This is signaled by a non-zero ground-state energy density, $E_0/N > 0$~[\onlinecite{Witten1981}], and the presence of gapless Nambu-Goldstone fermions~[\onlinecite{Salam1974}].

\subsection{Symmetries}

We now discuss the relevant symmetries of the model for arbitrary $g$. Additional symmetries appear at the infinite-coupling point, $g=0$, and will be discussed in Sec.~\ref{sec:inf-interacting}. The fermion parity $P = (-i)^{N/2} \prod_j \chi_j$ commutes with the Hamiltonian, which comprises only two- and four-fermion terms. The fermion parity also anticommutes with the supercharge $Q$. We thus have
\begin{align}
    [H, P] = 0 \quad , \quad \{Q, P\} = 0,
\end{align}
which implies that all (non-zero energy) eigenstates of $H$ are at least doubly degenerate~\footnote{In principle, the zero-energy manifold can have a different number of fermionic and bosonic eigenstates, because $Q \psi =0$ when $\psi$ is a zero-energy state. This difference is captured by the Witten index~[\onlinecite{Witten1982}]. On a lattice, the Witten index will be zero if the Hilbert space is a tensor product of local Hilbert spaces at each lattice site.} and form SUSY doublets $(\psi, Q\psi)$. The Hilbert space thus hosts two degenerate supersectors with fermion parities $P = \pm 1$, comprising ``bosonic" and ``fermionic" eigenstates, respectively. 

Throughout the paper, we consider periodic boundary conditions with $L_x$ and $L_y$ unit cells along the $x$ and $y$ directions shown in Fig.\,\ref{fig1}(a). This defines the translation operators $T_x$ and $T_y$, such that $[ H, T_{x,y}] = [Q, T_{x,y}] = 0$. Translations are subtle for systems of Majorana fermions, as they are only realized projectively. This results in the algebra of translation operators crucially depending on the system size~[\onlinecite{hsieh2016}]. For $L_x$ odd and $L_y$ even, we have 
\begin{equation}
   [P, T_x] = 0 \quad , \quad \{P, T_y \} = 0 \quad , \quad [T_x, T_y] = 0, \label{eq:Tx1}
\end{equation}
whereas for $L_x$ and $L_y$ both even,
\begin{equation}
    [P, T_x] = 0 \quad , \quad [P, T_y] = 0 \quad , \quad \{ T_x, T_y \} = 0. \label{eq:Tx2}
\end{equation}
The requirement of an even number of Majorana modes in the system prevents us from considering systems with $L_x$ and $L_y$ both odd. 

We thus have a set of mutually commuting operators, $H$, $P$ and $T_x$ which can be used to classify the eigenstates of $H$ in sectors. The fermion parity $P$ squares to 1 and thus has eigenvalues $\pm 1$. $T_x$ is unitary and respects $ \left(T_x\right)^{L_x} = 1$ -- it thus has eigenvalues $\re^{i\phi}$ where $\phi = 2\pi m/L_x$ with $m$ an integer. Given an eigenstate $\psi$, we can use the remaining symmetries $Q$ and $T_y$ to construct a degenerate subspace consisting of orthogonal states, as shown in Table \ref{tab:symmetries_finiteg}. Note that in the case with even $L_x$ and $L_y$, Eq.~(\ref{eq:Tx2}) results in an extra twofold degeneracy of the ground-state manifold, on top of the degeneracy expected from supersymmetry.

\renewcommand{\tabcolsep}{10 pt}
\begin{table}[]
\centering
\begin{tabular}{c|c|c|c}
  state        & $H$ & $P$ & $T_x$ \\
  \hline
  \hline
  $\ket{\psi_n}$     & $E_n$ & $1$   &  $\re^{i\phi}$ \\
  $T_y \ket{\psi_n}$ & $E_n$ & $\eta$   & $ -\eta \re^{i\phi}$ \\
  $Q \ket{\psi_n}$   & $E_n$ & $-1$  & $\re^{i\phi}$ \\
  $Q T_y \ket{\psi_n}$ & $E_n$ & $-\eta$  & $ -\eta \re^{i\phi}$ \\
\end{tabular}
 \caption{Structure of the degenerate subspaces of $H$ for even $L_x$ ($\eta = +1$) or odd $L_x$ ($\eta = - 1$). We take $L_y$ even in all cases. For odd $L_x$ the degeneracy is two-fold as mandated by the exact supersymmetry of the Hamiltonian. For even $L_y$ an additional two-fold symmetry emerges, with states distinguished by their eigenvalues under $T_x$.}
 \label{tab:symmetries_finiteg}
\end{table}

Finally, we can define a time-reversal operator $\Theta$ which simply sends $i \rightarrow -i$ and leaves all the $\chi_j$ invariant. This operator is constructed explicitly in the Appendix~\ref{app:symmetry} and combines complex conjugation $\mathcal{K}$ with a product of one half of the Majorana fermions. It acts as $\Theta H(g) \Theta^{-1} = H(-g)$, and thus the spectrum of $H(g)$ is identical to that of $H(-g)$. The model is also invariant under the point group symmetries of the kagome lattice, when compatible with the periodic boundary conditions. These symmetries are however not crucial to the present work.
\begin{figure}
    \centering
    \includegraphics[scale=0.45]{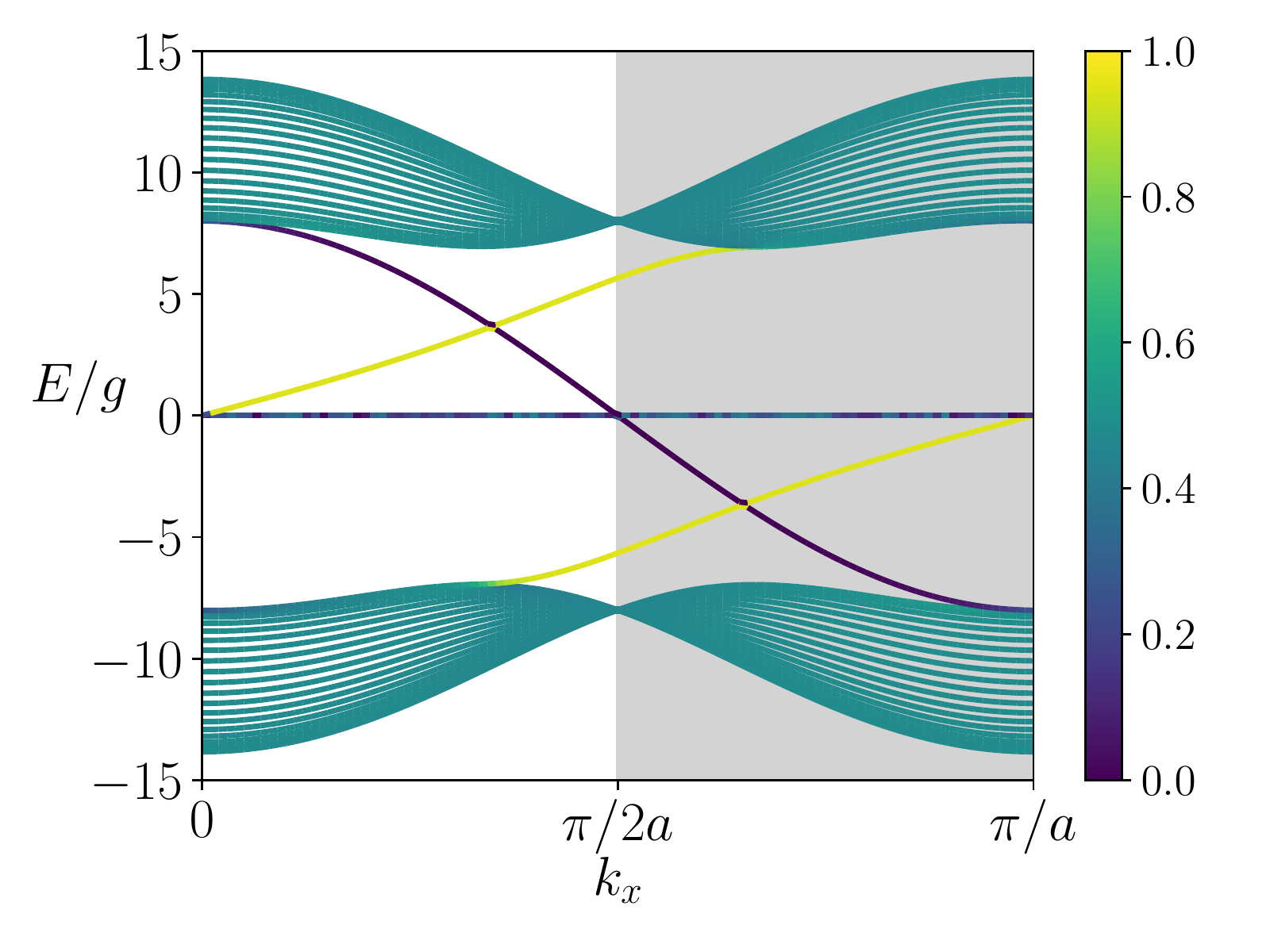}
    \caption{Spectrum of the non-interacting limit ($g=\infty$) of the model on a cylinder with open boundaries along the $y$ direction. The color scale represents the expectation value of the $\hat{y}$ position operator for each eigenstate. Chiral edge states connect the flat, zero-energy band with the positive and negative energy bands which carry Chern numbers $\mathcal{C} = \pm 1 $. Given the Majorana nature of the fermions, only half the Brillouin zone is physical, as shown by the shaded area.}
    \label{fig:spectrum}
\end{figure}
\section{Non-interacting limit}
\label{sec:non-interacting}
For $g = \infty$, the Hamiltonian becomes bilinear in Majorana fermions and can be written in momentum space as
\begin{equation}
H =  2 \sum_{k_x > 0, k_y} \Psi_{\bk}^\dg \mathcal{H}_\bk \Psi_{\bk},
\end{equation}
where $\Psi_\bk=(\chi^r_\bk,\chi^g_\bk,\chi^b_\bk)^T$ and $(r,g,b)$ color index denotes red, green and blue sublattice sites in Fig.\,\ref{fig1}(a). The Bloch Hamiltonian is a $3\times 3$ matrix in the sublattice space 
\begin{equation}
\mathcal{H}_\bk = 4gi
\begin{pmatrix}
 0 & \cos\frac{\bk\cdot\bd_y}{2} & -\cos\frac{\bk\cdot\bd_x}{2}\\
 & 0 & \cos\frac{\bk\cdot(\bd_x-\bd_y)}{2}\\
  & & 0
\end{pmatrix}+\mathrm{h.c.}
\end{equation}
with $\bd_x=(2a,0)$, $\bd_y=(a,\sqrt{3}a)$ and $a$ the nearest-neighbor distance. The energy spectrum 
\begin{align}
E_{\bk}
&= 0,\ \pm 4 g \sqrt{6 + 2 \cos 2 k_x a + 4 \cos k_x a \cos \sqrt{3} k_y a }
\end{align}
has a flat zero-energy band containing 1/3 of the states and two fully gapped bands with Chern number $\mathcal{C} = \pm 1$ (The flat band has $\mathcal{C} = 0$). This result differs slightly from the complex-fermion version of the model analyzed in Ref.~[\onlinecite{Bergman2008}, \onlinecite{Guo2009}], where the flat band sits at the top of the conduction band. In our case, the flat band sits at $E=0$ because $\mathcal{H}_\bk$ is  particle-hole symmetric and has an odd dimension.
In Fig.~\ref{fig:spectrum} we present the spectrum of $\mathcal{H}_\bk$ for a cylinder geometry with open boundary conditions along one direction. We observe two sets of chiral edge modes traversing the gap between the flat band and the two dispersive bands, confirming the assignment of Chern numbers given above. For $g=-\infty$ the spectrum is identical, but the assignment of Chern numbers of the dispersive bands is reversed.
\begin{figure*}
    \centering
    \includegraphics[width=\textwidth]{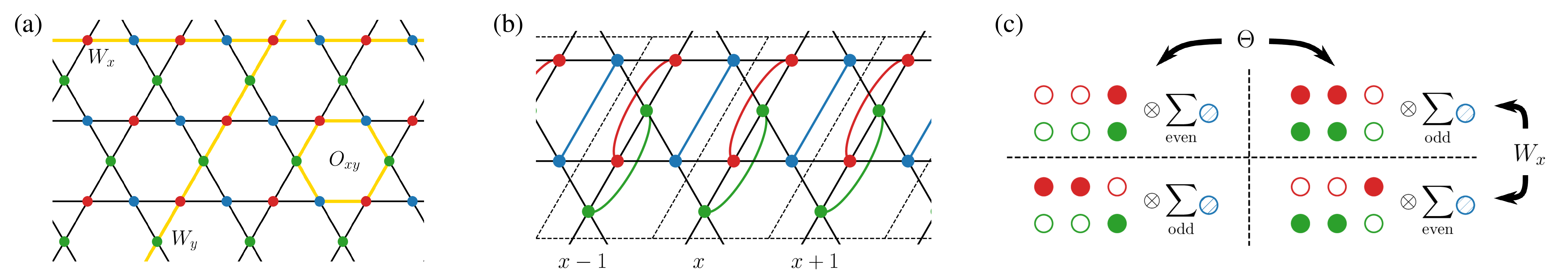}
    \caption{The infinite-coupling limit ($g=0$). (a) Two different types of operators commuting with $H$: local loops $\mathcal{O}_{xy}$ and topological loops $W_x$, $W_y$. (b) The pairing of Majorana fermions used to construct the complex fermion basis for thin tori ($L_y = 2$). (c) A cartoon for four exact ground states for $L_x=3$, $L_y=2$ in different sectors: fermion parity $P$ even/odd (left/right) and column parity $W_y$ even/odd (top/bottom). Filled/empty circles represent the presence/absence of a fermion on the corresponding site. These ground states cannot be connected by local operations. Acting with $W_x$ switches topological sector by flipping red and blue fermions, and acting with time-reversal $\Theta$ switches supersector by flipping all fermions. }
    \label{fig:loops}
\end{figure*}

\section{Infinite-coupling limit}
\label{sec:inf-interacting}

We now focus on the $g = 0$ limit. At that point the model has more symmetries. First, the Hamiltonian now commutes with time-reversal, $\Theta H(0) \Theta^{-1} = H(0)$. Next, we consider the product of all Majorana fermion operators on a given sublattice [which we label $\alpha = r, g, b$ as in Fig.\,\ref{fig1}(a)], 
\begin{equation}
P_\alpha = (-i)^{M/2} \prod_{j \in \alpha} \chi_j,
\end{equation}
where $M = N/3$ is the number of sites on each sublattice. The fermion parity operator $P$ is equal to $P_{r} P_{g} P_{b}$ (up to a global phase factor) and
\begin{equation}
    [H, P_\alpha] = 0 \quad , \quad \{ Q, P_\alpha\} = 0,
\end{equation}
such that individual color parity $P_\alpha$ is a good quantum number. The commutation relations with $T_x$ and $T_y$ also depend on system size, in a similar way as with $P$.

For $g=0$, the supercharge $Q$ only has triangular plaquettes. Therefore, any operator 
\begin{equation}
W_{\mathcal{L}} = \prod_{j, k \in \mathcal{L}} (i \chi_j \chi_k)
\end{equation}
taken along a closed path $\mathcal{L}$ 
touching $0$ or $2$ sites per triangular plaquette will commute with $Q$. These loops are of two fundamentally different types -- local loops comprising hexagons $\mathcal{O}_{xy}$ (or products of hexagons), and topological loops $W_x$, $W_y$ winding around the torus [see Fig.\,\ref{fig:loops}(a)]. Any local loop crosses the path of a topological loop at an even number of sites, and thus commutes with it. This means that the eigenvalues $\pm1$ of $W_x$, $W_y$ \emph{cannot} be changed by local operators commuting with $H$, and can be used to define topological sectors in the ground state manifold. Note that $\{ W_x, W_y \} = 0$, so that the eigenvalues of one loop define two topological sectors, whereas the other loop acts to switch between them. Such topological loops can be thought of as \emph{subsystem symmetries} as they act on sub-dimensional manifolds of the kagome lattice.

The hexagon operators $\mathcal{O}_{xy}$ are symmetries of the Hamiltonian, but do not form a mutually-commuting set. Indeed, neighboring hexagons anti-commute as they share a single site. Thus, the set of symmetries $\mathcal{O}_{xy}$ is responsible for a large degeneracy, as applying $\mathcal{O}_{xy}$ on an eigenstate of $H$ leads to an eigenstate with the same energy but different eigenvalues for the neighboring hexagons. Because the number of such hexagon operators scales linearly with the system size, we expect the induced degeneracy to scale exponentially with the system size. 

In the next subsection, we provide an explicit solution for the ground state manifold in the thin-torus limit which puts the above ideas on a more concrete footing.

\subsection{Exact solution for the ground state manifold}

The extra symmetries discussed above render the $g=0$ limit simpler than the generic model, but still seems intractable. The bowtie interaction terms do not commute, and we find that all ground states are entangled in real space, in contrast to known 1D SUSY Majorana models [\onlinecite{Fendley2018}, \onlinecite{Sannomiya2019}]. It turns out that thin-torus geometries -- systems of size $L_x \times 2$ with $L_x$ any integer -- are simpler because the real-space entanglement is limited to only one color (or sublattice). This allows us to construct the exact solution for the ground state manifold -- that is, to find all the zero-energy eigenstates of $H$. We outline the argument here and relegate technical aspects to Appendix~\ref{app:detail}. 

To set the stage, we pair same-colored Majorana fermions into complex fermions, as shown in Fig.\,\ref{fig:loops}(b). The solution then proceeds in two steps. First, we consider only the sector of the Hilbert space spanned by green and red fermions. We show that states $\ket{rg}$ where every column $x$ has the same fermion parity are \emph{annihilated} by the sum of green-red bowtie terms. These states are thus good building blocks to construct zero-energy eigenstates. Note that such column parities are equivalent to the loop operators $W_y$ defined above, and thus describe the two topological sectors (where \emph{all} $W_y$ have eigenvalues either $+1$ or $-1$). Each topological sector comprises $2^{L_x}$ states, as two configurations per column give rise to the same parity -- for example, see Fig.\,\ref{fig:loops}(c). 

We now turn our attention to the blue sector. Considering the remaining bowtie terms, we show that the problem reduces to a \emph{non-interacting} 1D chain described by Hamiltonian
\begin{align}
    H_{\text{blue}} = 4 \sum_x \left( \xi^r_x b_{x-1} b_x^\dagger + i  \xi^g_x  b_{x-1} b_x \right) + \mathrm{H.c.}.
    \label{eq:Hblue}
\end{align}
Here $\xi_x^\alpha = 2n_x^\alpha-1$ ($n_x^\alpha = 0,1$ are occupation numbers) are fixed to $\pm 1$ by the procedure described above and $b_x^\dagger$, $b_x$ are blue fermion operators. The two terms in Eq.~(\ref{eq:Hblue}) represent, respectively, nearest-neighbor hopping and pair creation/annihilation. The $\xi_x^\alpha$ factors can be consistently absorbed in a redefinition of the blue fermion operators, such that Eq.~(\ref{eq:Hblue}) becomes the celebrated Kitaev chain model with zero chemical potential [\onlinecite{Kitaev2001}]. In momentum space, the spectrum of $H_\text{blue}$ consists of two flat bands at energy $\pm4$, and its many-body ground state $\ket{b}$ has energy $-\frac{2N}{3}$. This ensures that
\begin{equation}
    H \left( \ket{rg} \otimes \ket{b} \right) = 0
\end{equation}
for each state $\ket{rg}$ constructed above (remembering the constant factor $+\frac{2N}{3}$ in Eq.~(\ref{eq:Hamiltonian})).

This systematic procedure allows us to construct a zero-energy manifold of size $\Omega = 2^{L_x+1}$, as shown in Table~\ref{tab:deg}. The ground state wavefunctions are quite peculiar -- they are product states in the red and green sectors, whereas the blue sector comprises an equal weight superposition of all configurations of given color parity [see Fig.\,\ref{fig:loops}(c)]. Note that the fermion parity (or supersector) of a state is \emph{independent} of its topological sector (the column parities). Thus, the ground state manifold splits into $4$ sectors which cannot be connected by local, commuting operators \footnote{Local but \emph{non-commuting} operators can have non-trivial matrix elements between the two supersectors, but not between the two topological sectors. Also, no local operator can distinguish between topological sectors or supersectors, which are defined by non-local fermion parities.}. Within each sector, all ground states can be connected by local operators comprising products of hexagons $\mathcal{O}_{xy}$. To check that our method allows to construct the entire zero-energy manifold, we numerically obtained the ground states using exact diagonalization (up to system sizes $8 \times 2$) and verified the one-to-one correspondence with the analytical result \footnote{The $2 \times 2$ case is special because $H$ is invariant under $R_3$ rotations, which are compatible with the periodic boundary conditions. There is an extra degeneracy of 3 in the numerical ground states because our basis choice breaks this rotational symmetry. The missing ground states can be constructed by rotating the basis by $\pm 2\pi/3$.}. 

\renewcommand{\tabcolsep}{5.4pt}
\begin{table}[]
\centering
\begin{tabular}{c||c|c|c|c|c|c|c|c}
$N$      & 18 & 24 & 30 & 36 & 36 & 42 & 48 & 54 \\
\hline
$L_x, L_y$ & $3,2$  & $4,2$  & $5,2$  & $6,2$  & $3,4$ & $7,2$ & $8,2$ & $3,6$ \\
\hline
\hline
$\Omega (g = 0)$      &  16 & 32 & 64 & 128 & 128 & 256 & 512 & 512 \\
$\Omega (g \neq 0)$ & 4  & 8  & 4  & 8   & 4   & 4 & n/a & n/a \\
    \end{tabular}
    \caption{Ground state degeneracy $\Omega$ for small systems obtained from exact diagonalization. 
    For $N >42$, only the $g=0$ limit is accessible because of its extra symmetries.}
    \label{tab:deg}
\centering
\end{table}

Our solution does not easily generalize to the 2D limit, as the entanglement pattern is more complicated and involves all colors. Numerically, we checked that systems of size $3\times 4$ and $3\times 6$ also have degenerate ground states with $\Omega= 128$ and $512$ respectively. This indicates that the exponential scaling is a truly 2D property, independent of the thin-torus geometry. The topological sectors are also expected to survive, as no local operator can change the eigenvalue of all loops $W_y$. However, we find a small but non-zero ground state energy for the largest system size amenable to exact diagonalization, $3\times 6$ ($N=54$). Whether or not SUSY is broken spontaneously at $g=0$ in the 2D limit is thus still an open question.

\subsection{Excitations}

\begin{figure}
    \centering
    \includegraphics[width=0.38\textwidth]{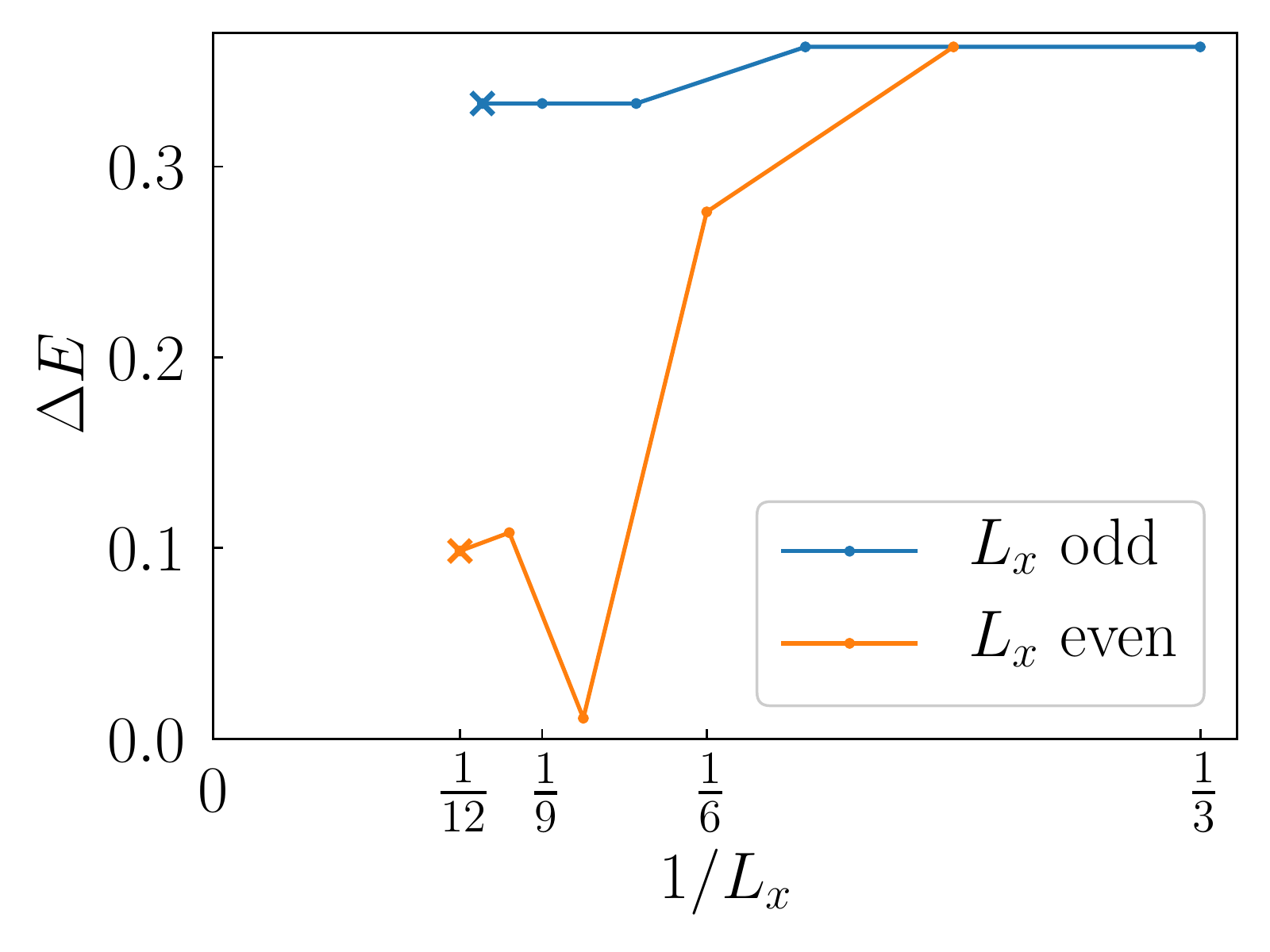}
    \caption{The scaling of the energy gap $\Delta E$ to the first excited state in the $g=0$ limit. Dots represent reliable energies, whereas crosses are an upper bound (see the Appendix~\ref{app:numerics} for details on numerics for very large systems).}
    \label{fig:gap}
\end{figure}

This exact solution unfortunately does not inform us on the structure of excitations, which represents a much harder problem. For even $L_x$, numerical results suggest that the model might be gapless in the thermodynamic limit (see Fig.\,\ref{fig:gap}). We find that the lowest excited state manifold is also exponentially degenerate and comprises states with \emph{alternating} column parities for large enough $L_x$, in stark contrast to the ground state manifold. Note that such excited states cannot be locally created from a ground state -- one needs to apply an operator (of size scaling at least linearly with $L_x$) which anti-commutes with one half of the loops $W_y$. For odd $L_x$, alternating column parity states are impossible. Whether the theory is gapped or gapless is unclear (see Fig.\,\ref{fig:gap}) and will require more sophisticated numerical methods to settle.

\section{SUSY breaking at finite coupling}
\label{sec:finiteg}

When $g \neq 0$, the color parities $P_\alpha$ and loops are no longer symmetries, and thus the exponential degeneracy collapses to a fixed number independent of system size ($8$ for even $L_x$ and $4$ for odd $L_x$), as shown in Table~\ref{tab:deg}. Note that there is an extra degeneracy of $2$ compared to the minimal degeneracy enforced by SUSY and translation symmetry (c.f. Table~\ref{tab:symmetries_finiteg}). This is likely a consequence of the point group symmetries which were not included in our analysis.

\begin{figure*}
    \centering
    \includegraphics[width=0.85\textwidth]{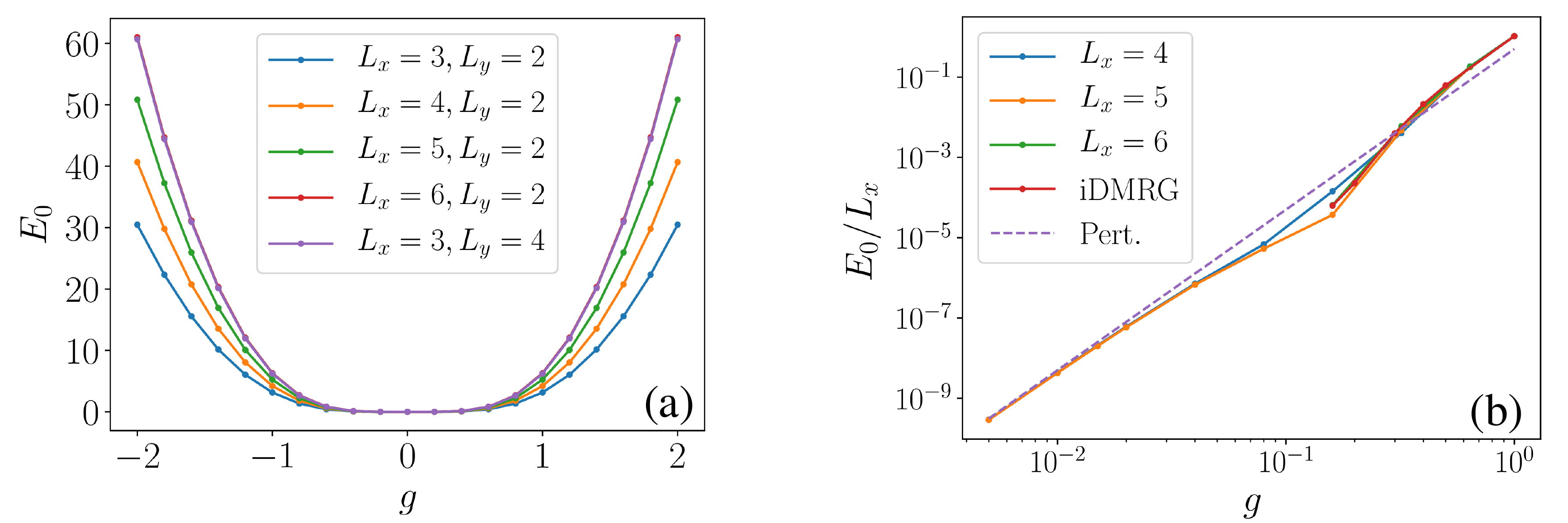}
    \caption{(a) The ground state energy as a function of $g$ for different system sizes. (b) Comparison between the ground state energy density for small $g$, obtained using ED and iDMRG, and the scaling expected from fourth-order perturbation theory.}
    \label{fig:E0}
\end{figure*}

Using exact numerical diagonalization for system sizes up to $N=42$, we find that the ground state energy $E_0$ is nonzero for finite $g$, indicating spontaneously-broken SUSY. The breaking of SUSY near $g=0$ is however very weak, with $E_0$ scaling as $g^4$ [see Fig.\,\ref{fig:E0}]. The question of whether SUSY is recovered in the thermodynamic limit -- that is, if $E_0/N \rightarrow 0$ when $N \rightarrow \infty$, as seen in Refs.~[\onlinecite{Fendley2018}, \onlinecite{Sannomiya2019}], is difficult to answer unequivocally with numerical exact diagonalization. To resolve this question, we adopt two complementary approaches: i) infinite density-matrix renormalization group (iDMRG) calculations for cylinders of width $L_y = 2$, and ii) a numerical Schrieffer-Wolff degenerate perturbation theory calculation~[\onlinecite{Bravyi2011}].  Note that an analytical argument (see Appendix~\ref{app:groundstate}) implies that, on general grounds, SUSY is necessarily broken (in the thermodynamic limit) for $g \geq 1.6$. However, this argument cannot distinguish what happens for $g < 1.6$.

Using iDMRG, we obtain the ground state energy density of the model in the thin-torus geometry, $L_y = 2$, directly in the thermodynamic limit ($L_x \rightarrow \infty$). We use a supercell comprising $N=12$ complex fermions (corresponding to $L_x=4$) and verify convergence as a function of a bond dimension $\chi$ up to $\chi=2000$. This task is complicated by the fact that the energy gap of the system is decreasing rapidly as $g \rightarrow 0$, which reflects the fact that the ground state at $g=0$ is exponentially degenerate. Converging the iDMRG simulations is therefore very delicate for small $g$. We were only able to obtain reliable results for $g \geq 0.16$ when using a maximal bond dimension $\chi=2000$. For these cases we find a non-zero ground state energy density consistent with exact diagonalization results, as shown in Fig.~\ref{fig:E0}(b).

The fact that $E_0 \sim g^4$ (in exact diagonalization) for small $g$ suggests that a fourth-order perturbation theory around the exponentially-degenerate manifold at $g=0$ should capture the SUSY breaking. Using a numerical algorithm implementing the Schrieffer-Wolff method~[\onlinecite{Bravyi2011}], we find that this is indeed the case, as summarized in Table~\ref{tab:pert_theory}. For all accessible system sizes ($L_x=3,4,5,6$ and  $L_y=2$), the first and third-order corrections to the ground state energy vanish identically. The second-order correction $E_0^{(2)} = -N g^2$ exactly cancels out the last term in the Hamiltonian, Eq.~(\ref{eq:Hamiltonian}), such that the ground state energy remains zero up to third-order in perturbation theory. This explains the weak SUSY breaking observed in exact diagonalization. To fourth order, we find that
\begin{equation}
    \frac{ E^{(4)}_0}{N} = \frac{g^4}{12},
\end{equation}
corresponding to the spontaneous breaking of SUSY in the thermodynamic limit. This result agrees quantitatively with exact diagonalization results for various system sizes, as shown in Fig.~\ref{fig:E0}(b). There is however no splitting of the exponentially-degenerate ground state manifold to fourth order. This splitting occurs in fifth (and higher) orders, as shown in Table \ref{tab:pert_theory}.

\begin{table}[]
\centering
\begin{tabular}{ c | c c c c c | c} 
  $L_x$ & $c_1$ &  $c_2$ &  $c_3$ &  $c_4$ &  $c_5$ & $\Delta_5$\\ 
  \hline
  3 & 0 & $-18$ & 0 & $1.5$ & $ \pm 126.2, \pm 31.8 $ & 94.4\\ 
  4 & 0 & $-24$ & 0 & $2$ & $ \pm 24 $ & 48 \\ 
  5 & 0 & $-30$ & 0 & $2.5$ & $ \pm 36.9, \pm 8.7 $ & 28.2\\ 
  6 & 0 & $-36$ & 0 & $3$ & n/a & n/a\\ 
  \hline
\end{tabular}
\centering
\caption{Results of the degenerate perturbation theory around the infinite-coupling limit, $g=0$, in the ground state manifold. We list the $n$th-order corrections $E_0^{(n)} = c_n g^n$ to the ground state energy for systems in thin-torus geometries ($L_y=2$). To fifth-order the ground state manifold splits, giving rise to an induced gap $E_{gap}^{(5)} = \Delta_5 g^5$.}
\label{tab:pert_theory}
\end{table}

The spontaneous breaking of SUSY at $g \neq 0$ is expected to give rise to Nambu-Goldstone fermions~[\onlinecite{Volkov1973},\onlinecite{Salam1974}] that are gapless in the thermodynamic limit. Their identification is beyond the scope of the numerical methods employed here, but we note that the gap to the first excited state induced to fifth-order (the lowest non-trivial order splitting the ground state manifold)  decreases rapidly with system size -- providing a suggestive hint. Confirming the presence of Nambu-Goldstone fermions using more sophisticated numerical methods, and understanding their dispersion relation (which is cubic for the 1D SUSY Majorana case~[\onlinecite{Sannomiya2019}]) could be interesting directions for future work.

\section{Outlook}

In this work, we constructed a supersymmetric model of interacting Majorana fermions on the kagome lattice. The infinite-coupling limit ($g=0$) of the model admits an unbroken supersymmetric point with some peculiar features. The ground state manifold is exponential in system size, and splits into four sectors that cannot be connected by local, commuting operators: two supersymmetric sectors defined by the global fermionic parity, and two topological sectors defined by the fermionic parities along one-dimensional submanifolds of the system. In the limit of thin-torus geometries, we presented an exact analytical solution for the entire zero-energy ground state manifold. This solution relies on the observation that ground state wavefunctions with a specific pattern of entanglement (limited to one sublattice) can be found. Unfortunately, we have not been able to generalize this solution to the truly two-dimensional limit. Whereas the exponential scaling of the ground state manifold and the presence of topological sectors are expected to survive in the two-dimensional limit, numerical evidence about whether or not SUSY is spontaneously broken at $g=0$ is not available at the moment. 

For finite-coupling, degenerate perturbation theory and iDMRG calculations indicate that SUSY is spontaneously broken, albeit very weakly, with a ground state energy density $E_0/N \sim g^4$. This spontaneous breaking of a ``fermionic symmetry" should be accompanied by gapless Nambu-Goldstone fermionic modes for $g\neq0$, although numerical evidence for this is scarce.

It is worth noting that the results of Ref.~[\onlinecite{hsieh2016}] imply that all translation-symmetric lattice models with an odd number of Majorana fermions per unit cell exhibit an $\mathcal{N} = 2$ SUSY. Here, we constructed explicitly a model with $\mathcal{N} = 1$ SUSY which clearly involves physics beyond the construction employed in Ref.~[\onlinecite{hsieh2016}].

Many open questions remain -- for example, what is the nature of the spontaneously-broken SUSY phase realized for finite $g$? This is not obvious since the non-interacting limit has a completely-flat band at zero-energy, and thus even weak interactions could have an important effect. A more detailed analysis of the excitations of the model in the infinite-coupling is also called for, and could reveal interesting structures owing to the presence of topological sectors. Furthermore, reliable large-scale numerics are needed to determine the fate of SUSY far from the thin-torus geometries, and whether the model at $g=0$ is gapped or gapless in the thermodynamic limit. 

Our construction can be applied more generally to any lattice comprising corner-sharing triangles, such as triangular in 2D and hyperkagome or pyrochlore in 3D. The resulting Hamiltonians in these cases are expected to be more complex. Nevertheless, it is tempting to wonder if an unbroken supersymmetric \emph{phase} (extending over a finite region of parameter space), similar to those seen in recently proposed 1D models~[\onlinecite{Fendley2018}, \onlinecite{Sannomiya2019}], could exist in such higher-dimensional interacting Majorana systems.


\section{Acknowledgments}

We thank Ian Affleck, Stephan Plugge, Sharmistha Sahoo, and Tarun Tummuru for helpful discussions. The work described in this article was supported by NSERC and CIfAR. C. L. and \'E. L.-H. were also supported by the QuEST scholarship at the University of British Columbia. The iDMRG calculations were performed using the ITensor library [\onlinecite{ITensor}].

C. L. and \'E. L.-H. contributed equally to this work.

\appendix

\section{Further discussion of symmetries}
\label{app:symmetry}

\emph{Time-reversal symmetry.} -- Here we construct the time-reversal operator $\Theta$, such that $\Theta \chi_k \Theta^{-1} = \chi_k$ and $\Theta i \Theta^{-1} = -i$. Therefore any real Hamiltonian built out of $\chi_j$ will be invariant under $\Theta$. For systems with an even number $N/2$ of complex fermions, we choose
\begin{align}
    \Theta = \left( \prod_{j=1}^{N/2} \chi_{2j} \right) \mathcal{K},
\end{align}
where $\mathcal{K}$ is the complex conjugation operation ($\mathcal{K}^2 = 1$), and we use the convention where
\begin{align}
\chi_{2j} = i (c_j - c_j^\dagger) \quad , \quad \chi_{2j-1} = c_j^\dagger + c_j.
\end{align}
With this definition it is easy to check that $[\Theta, \chi_k]$ = 0. If $k$ is odd, $\chi_k$ commutes with $\mathcal{K}$ and with the product (which does not contain $\chi_k$). If $k$ is even, acting with $\mathcal{K}$ picks up a minus sign, and another minus sign comes from anti-commuting with the product (which now contains $\chi_k$). Thus $\left[ \Theta , \chi_k \right] = 0 $ and $\Theta \chi_k \Theta^{-1} = \chi_k$.
For systems with an odd number $N/2$ of complex fermions, we instead choose a product including the other half of the Majorana fermions,
\begin{align}
    \Theta = \left( \prod_{j=1}^{N/2} \chi_{2j-1} \right) \mathcal{K}.
\end{align}
By the same reasoning as above, we find that $\left[ \Theta , \chi_k \right] = 0 $.

\emph{Translation operators.} -- From the symmetry between rows and columns, it suffices to consider the cases with even $L_y$. In this case we can construct explicitly the translation operators from the Majorana operators as
\begin{equation}
T_x \propto\prod_{\alpha=\mathrm{r,g,b}}\prod_{j=1}^{L_y}\prod_{i=L_x}^2(\chi_{1,j,\alpha}-\chi_{i,j,\alpha}).
\end{equation}
Note that the order can not be changed arbitrarily. It is straightforward to check that $T_x$ is unitary and
\begin{equation}
T_x\chi_{i,j,\alpha}T_x^{-1}=\chi_{i+1 (\mathrm{mod}\ L_x),j,\alpha}.
\end{equation}
Combining this with the requirement that
\begin{equation}
T_y\chi_{i,j,\alpha}T_y^{-1}=\chi_{i,j+1 (\mathrm{mod}\ L_y),\alpha},
\end{equation}
it is straightforward to check Eqs. (\ref{eq:Tx1}) and (\ref{eq:Tx2}).

\emph{Hexagonal plaquettes at $g=0$.} -- One interesting feature of the model at $g=0$ is that we can identify a set of mutually commuting hexagonal plaquette operators $\mathcal{O}_{xy}$,
\begin{equation}
    \mathcal{O}_{xy} = i \prod_{j \in \hexagon} \chi_j,
\end{equation}
which also commute with the Hamiltonian. Note that neighboring hexagonal plaquettes anti-commute on the kagome lattice as they only share one site, but one can still pick one third of the hexagons to form a set of mutually commuting operators (compatible with the ``Kekul{\'e}" tiling). There is one such hexagon for every three unit cells, that is, $N/9$ hexagons.

\begin{figure}
    \centering
    \includegraphics[scale=0.26]{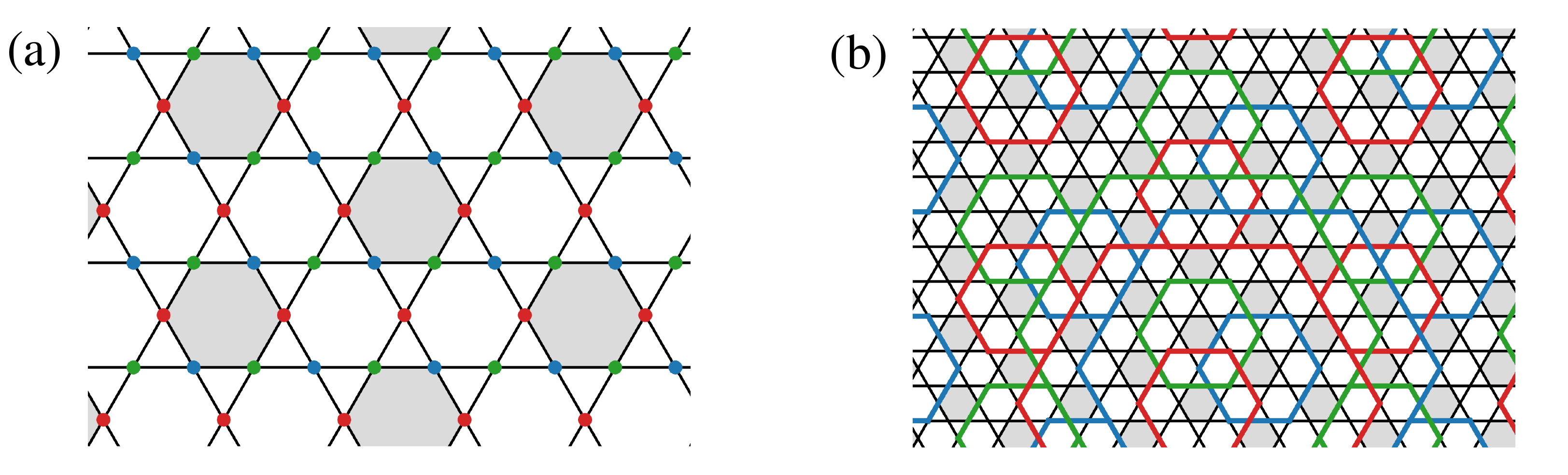}
    \caption{(a) Mutually commuting hexagons. (b) The infinite order Kekul{\'e} tiling.
    \label{fig:infinite_kekule}}
\end{figure}

One can extend this construction to a coarse-grained kagome lattice with side length $3^n a$ (where $a$ is the nearest-neighbor distance), as shown in Fig.\,\ref{fig:infinite_kekule}(a) and (b). That is, for each $n$, we can construct similar hexagons at a larger scale. Importantly, this new set of operators commute with every operator from the previous steps (i.e., shorter lengthscales). The maximal number of such mutually-commuting operators is $1/3$ of the operators at the previous step $n-1$. We thus sum the geometric series,
\begin{equation}
    \frac{N}{9}\left(1 + \frac{1}{3} + \frac{1}{3^2} + ... \right) = \frac{N}{6}.
\end{equation}
This set of conserved quantities spans a Hilbert space of at most $2^{N/6}$, whereas the total Hilbert space dimension scales as $2^{N/2}$. This is therefore not enough to afford us with an exact solution in the 2D limit. Also note that not all configurations spanned by these hexagons are allowed in the ground state -- for example, it is not possible to flip the eigenvalue of single hexagon without leaving the ground state manifold. 

\section{Details of exact solution}
\label{app:detail}

\begin{figure}
    \centering
    \includegraphics[scale=0.45]{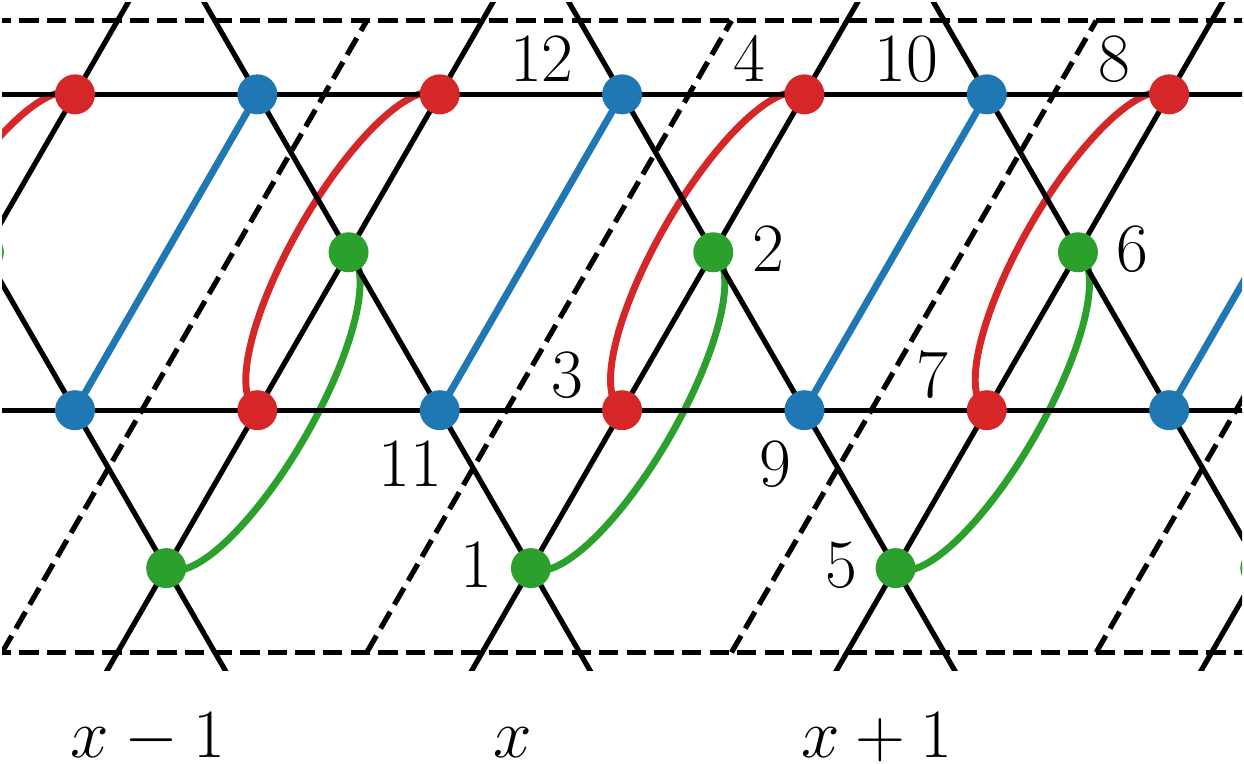}
    \caption{The pairing of Majorana fermions used to construct the complex fermion basis.}
    \label{fig:basis}
\end{figure}

Here we present details of the exact solution presented in the main text for system sizes $L_x\times 2$. We first fix the complex fermion basis for the problem using the pairing shown in Fig.\,\ref{fig:basis}. This basis calls for partitioning the solution, by looking first at individual columns labeled by their position $x$. We also consider separately the bowtie terms which contain blue sites and the bowtie terms which do not.

\emph{No blue fermions} -- There are two green-red bowties per column (let us call them $B^x_1$ and $B^x_2$ for the column labeled by $x$), and any such pair of bowties act on the same four complex fermions. Those are defined, following the basis in Fig.~\ref{fig:basis}, as
\begin{align}
    g_x     &= \frac{1}{2} \left( \chi_1 - i \chi_2 \right) \quad , \quad
    g_{x+1} = \frac{1}{2} \left( \chi_5 - i \chi_6 \right),  \nonumber \\
    r_x     &= \frac{1}{2} \left( \chi_3 - i \chi_4 \right) \quad , \quad
    r_{x+1} = \frac{1}{2} \left( \chi_7 - i \chi_8 \right).
    \label{eq:complexfermion_definition}
\end{align}
Now we re-express the red-green bowties $B^x_{1,2}$ in terms of complex fermions by inverting the definition, Eq.~(\ref{eq:complexfermion_definition}):
\begin{equation}
    \begin{split}
    B^x_1 &= \chi_3 \chi_2 \chi_5 \chi_7  \\
    &= -i ( r_x^\dagger + r_x  ) ( g_x^\dagger - g_x ) ( g_{x+1}^\dagger + g_{x+1} ) ( r_{x+1}^\dagger + r_{x+1} ),  \\
    B^x_2 &= \chi_4 \chi_1 \chi_6 \chi_8 \\
    &=  i( r_x^\dagger - r_x  ) ( g_x^\dagger + g_x ) ( g_{x+1}^\dagger - g_{x+1} ) ( r_{x+1}^\dagger - r_{x+1} ). 
    \end{split}
\end{equation}
Adding both terms, we are left with 
\begin{equation}
    \begin{split}
    B^x_1 + B^x_2
    &= 2 i( r_x^\dagger g_x g_{x+1} r_{x+1} - r_x g_x^\dagger g_{x+1} r_{x+1}  \\
    &+ r_x g_x g_{x+1}^\dagger r_{x+1} + r_x g_x g_{x+1}  r_{x+1}^\dagger   ) + H.c.
    \end{split}
\end{equation}
It follows that all states with an even occupation number (among the four fermions considered) are annihilated by $B^x_1 + B^x_2$. These are thus good building blocks to construct zero-energy eigenstates. Note that either both columns have an even number of fermions, or both columns have an odd number fermions.
Now, we need to combine this solution space with that of the next column, where $B_1^{x+1} + B_2^{x+1}$ annihilates states with an even number of fermions on the adjacent columns $x+1$ and $x+2$. Carrying this procedure iteratively, we see that in the ground state \emph{all} columns have the same fermion parity. This is the condition discussed in the main text.

\emph{Blue fermion terms} -- Now, we deal with terms involving blue fermions. Let us define our basis. We re-use the fermions $g_x$ and $r_x$, and introduce the blue fermions
\begin{align}
b_x     &= \frac{1}{2} ( \chi_9 - i\chi_{10} ) \quad , \quad
b_{x-1} = \frac{1}{2} ( \chi_{11} - i \chi_{12} ).
\label{eq:complexfermion_definition_blue}
\end{align}
There are four blue-containing bowties per column $x$ -- two red-blue bowties labeled $A^x_1$ and $A^x_2$ and two green-blue bowties labeled $C^x_1$ and $C^x_2$. In the complex fermion basis, using the definitions Eqs.~(\ref{eq:complexfermion_definition}) and (\ref{eq:complexfermion_definition_blue}), those read:
\begin{align}
    A^x_1 &= \chi_9 \chi_3 \chi_4 \chi_{12}  = -\xi^r_x (b_{x-1}^\dagger - b_{x-1})  ( b_x^\dagger + b_x ),  \nonumber \\
    A^x_2 &= \chi_{11} \chi_3 \chi_4 \chi_{10}   = \xi^r_x ( b_{x-1}^\dagger + b_{x-1}) ( b_x^\dagger - b_x ),  \nonumber \\
    C^x_1 &= \chi_1 \chi_{10} \chi_{12} \chi_2 = i \xi^g_x ( b_{x-1}^\dagger - b_{x-1}) ( b_x^\dagger - b_x ),  \nonumber \\
    C^x_2 &= \chi_1 \chi_{11} \chi_{9} \chi_2  = i \xi^g_x (b_{x-1}^\dagger + b_{x-1} ) ( b_x^\dagger + b_x ),
\end{align}
where we simplified the expressions using $\xi^r_x = i \chi_4 \chi_3$ and $\xi^g_x = i \chi_2 \chi_1$, where $\xi=-1$ ($\xi=1$) represents an empty (occupied) fermionic state. Note that the bowties act trivially on the green and red parts of the wavefunctions. This is a crucial property of $L_x \times 2$ systems which render this exact solution possible. Again, summing bowties of the same type, we get a simple form
\begin{align}
    A^x_1 + A^x_2 &= 2 \xi^r_x (b_{x-1} b_x^\dagger - b_{x-1}^\dagger b_x ),\nonumber \\
    C^x_1 + C^x_2 &= 2 i  \xi^g_x ( b_{x-1} b_x + b_{x-1}^\dagger b_x^\dagger ).
    \label{eq:suppl_bluebowties}
\end{align}
Note that $A^x_1 + A^x_2$ hops blue fermions between neighboring blue fermion sites,
whereas $C^x_1 + C^x_2$ creates or annihilates pairs of blue fermions on neighboring site. The values of $\xi^r_x$ and $\xi^g_x$ are fixed for all $x$ by our choice of red-green product state $\ket{rg}$ annihilated by all $B^x_1 + B^x_2$ (as discussed in the main text, there are $2^{L_x+1}$ such choices). Combining the two terms in Eq.~(\ref{eq:suppl_bluebowties}), we obtain Eq.~(\ref{eq:Hblue}) in the main text, describing a one-dimensional chain of blue fermions analogous to the Kitaev chain model. As an example, we give the explicit form of the ground state wavefunctions for system sizes $2 \times 2$ and $3 \times 2$ in the next section.

\section{Numerical solution}
\label{app:numerics}

For $2 \times 2$, we obtain the following 8 wavefunctions which respect the previous construction. In the even parity sector, we have
\begin{equation}
\begin{split}
&\ket{01}_r \ket{01}_g \left( \ket{00} - i\ket{11} \right)_b , \\
&\ket{10}_r \ket{10}_g \left( \ket{00} + i\ket{11} \right)_b,\\
&\ket{10}_r \ket{01}_g \left( \ket{00} - i\ket{11} \right)_b, \\
&\ket{01}_r \ket{10}_g \left(\ket{00} + i\ket{11} \right)_b.
\end{split}
\end{equation}
and in the odd parity sector,
\begin{equation}
\begin{split}
& \ket{00}_r \ket{00}_g \left( \ket{01} - \ket{10} \right)_b ,\\
& \ket{11}_r \ket{11}_g \left(\ket{01} + \ket{10} \right)_b, \\
& \ket{11}_r \ket{00}_g \left( \ket{01} + \ket{10} \right)_b,\\
&\ket{00}_r \ket{11}_g \left( \ket{01} - \ket{10} \right)_b.
\end{split}
\end{equation}
Within a supersector, the states are further distinguished by the eigenvalues under topological loops $W_y$: the first two states have even column parity and the last two have odd column parity. Finally, within such topological sectors, the states are always related by a local operation: an hexagon which flips all 4 red and green fermions.

Note that the $2 \times 2$ Hamiltonian is symmetric under $R_3$ rotations, whereas the basis choice is not. Therefore, the ground states with blue entanglement are not special but a consequence of the basis choice -- we should have the same structure, had we chosen a basis with red or green as a special color. These states are seen in the other color parity sectors, and have more complicated entanglement in the present basis. This generates a three-fold degeneracy bringing the number of ground states to 24 (whereas the naive counting presented above yields 8).

\begin{figure*}
    \centering
    \includegraphics[width=0.95\textwidth]{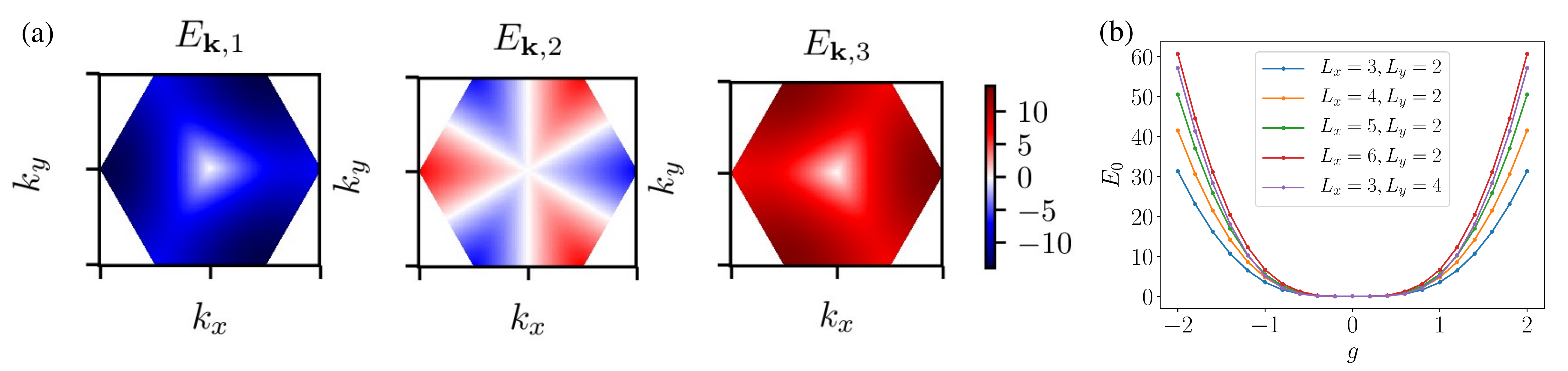}
    \caption{The generalized model [Eq.~(\ref{eq:H_generalized})]. (a) Spectrum for the three bands in the non-interacting limit ($g=\infty$). (b) The ground state energy as a function of $g$ for different system sizes.}
    \label{fig:spectrum_generalized}
\end{figure*}

For $2 \times 3$ the Hamiltonian is not rotation-symmetric, and thus blue is indeed special to start with. We do not have a three-fold degeneracy in the ground state manifold, and we observe that \emph{all} 16 ground states expected on the basis of the above construction are present:
\begin{equation}
\begin{split}
& \text{$P$ even , $W_y$ even} \\
&\ket{000}_r\ket{000}_g(\ket{000}+i\ket{011}-i\ket{101}+i\ket{110})_b,\\
&\ket{011}_r\ket{011}_g(\ket{000}-i\ket{011}-i\ket{101}-i\ket{110})_b,\\
&\ket{110}_r\ket{110}_g(\ket{000}+i\ket{011}+i\ket{101}-i\ket{110})_b,\\
&\ket{101}_r\ket{101}_g(\ket{000}-i\ket{011}+i\ket{101}+i\ket{110})_b,
\end{split}
\end{equation}
\begin{equation}
\begin{split}
& \text{$P$ even, $W_y$ odd} \\
&\ket{010}_r\ket{101}_g(\ket{001}-\ket{010}-\ket{100}+i\ket{111})_b,\\
&\ket{001}_r\ket{110}_g(\ket{001}+\ket{010}-\ket{100}-i\ket{111})_b,\\
&\ket{100}_r\ket{011}_g(\ket{001}-\ket{010}+\ket{100}-i\ket{111})_b,\\
&\ket{111}_r\ket{000}_g(\ket{001}+\ket{010}+\ket{100}+i\ket{111})_b,
\end{split}
\end{equation}
\begin{equation}
\begin{split}
& \text{$P$ odd, $W_y$ odd} \\
&\ket{011}_r\ket{100}_g(\ket{000}+i\ket{011}+i\ket{101}+i\ket{110})_b,\\
&\ket{000}_r\ket{111}_g(\ket{000}-i\ket{011}+i\ket{101}-i\ket{110})_b,\\
&\ket{110}_r\ket{001}_g(\ket{000}-i\ket{011}-i\ket{101}+i\ket{110})_b,\\
&\ket{101}_r\ket{010}_g(\ket{000}+i\ket{011}-i\ket{101}-i\ket{110})_b,
\end{split}
\end{equation}
\begin{equation}
\begin{split}
& \text{$P$ odd, $W_y$ even} \\
&\ket{100}_r\ket{100}_g(\ket{001}-\ket{010}+\ket{100}+i\ket{111})_b,\\
&\ket{111}_r\ket{111}_g(\ket{001}+\ket{010}+\ket{100}-i\ket{111})_b,\\
&\ket{010}_r\ket{010}_g(\ket{001}-\ket{010}-\ket{100}-i\ket{111})_b,\\
&\ket{001}_r\ket{001}_g(\ket{001}+\ket{010}-\ket{100}+i\ket{111})_b.\\
\end{split}
\end{equation}
Here the color parities are locked with the column parity. This is because, for odd system sizes, the loops $W_x$ change \emph{both} the column parities and red and blue color parities. Within each sector, all states can be related by local operations -- any three hexagons, which flip the parities of adjacent columns.

Here we briefly comment on the numerical methods for $g=0$. In order to approach the large system sizes ($N>42$), we construct a block Hamiltonian for each sector with definite column and color parities. For system sizes $L_x \times 2$, the dimension of each block scales as $2^{2L_x-2}$ while the number of blocks scales as $2^{L_x+2}$. This is a significant improvement over the full Hilbert space dimension $2^{N/2} = 2^{3L_x}$. The results in Fig.\,4(c) up to $2\times 10$ come from a thorough search of each sector. For $2\times 11$ ($N=66$) and $2\times 12$ ($N=72$), such a search is too expensive to complete. The results come from a search over a subset of the sectors chosen as a natural extrapolation of the first excited state manifold obtained for smaller systems. They thus give an upper bound for the gap energy, which is marked by a cross to emphasize the difference.

\section{Ground state energy in the non-interacting limit and SUSY breaking}
\label{app:groundstate}

We have, in the non-interacting limit,
\begin{equation}
    H = 2 i g \sum_{\langle\mathbf{r}, \mathbf{r'}\rangle} \chi_\alpha(\mathbf{r}) \chi_
\beta(\mathbf{r'}) 
\end{equation}
where the sum is over nearest-neighbors, and we label the sites according to their unit cell $\br, \br'$ and sublattice index $\alpha, \beta$. We Fourier-transform this expression using
\begin{align}
    \chi_\alpha( \br ) = \sqrt{\frac{2}{n}} \sum_\bk \chi^\alpha_\bk \re^{i \bk \cdot ( \br + \br_\alpha) }
\end{align}
where $\br_\alpha$ is the sublattice vector and $n=N/3$ is the number of unit cells. We then get $H = \sum_\bk \Psi_{-\bk} \mathcal{H}_\bk \Psi_\bk$ where $ \Psi_\bk = (\chi^r_\bk, \chi^g_\bk, \chi^b_\bk) $ and
\begin{align}
    \mathcal{H}_\bk = 4 i g \begin{pmatrix}
     0 & \cos \frac{\bk \cdot \bd_y}{2} & - \cos \frac{\bk \cdot \bd_x}{2} \\
     -\cos \frac{\bk \cdot \bd_y}{2} & 0 & \cos \frac{\bk \cdot (\bd_x - \bd_y) }{2} \\ \cos \frac{\bk \cdot \bd_x}{2} & -\cos \frac{\bk \cdot (\bd_x - \bd_y) }{2} & 0 
    \end{pmatrix}
\end{align}
For Majorana operators, $\{ \chi^\alpha_\bk, \chi^\beta_{\bk'} \}= \delta_{\bk, -\bk'} \delta_{\alpha \beta}$ and $(\chi^\alpha_\bk)^\dagger = \chi^\alpha_{-\bk}$, also note that $\mathcal{H}_\bk = \mathcal{H}_{-\bk}$. Therefore, we can rewrite $H$ using a sum over only one half of the Brillouin zone:
\begin{align}
    H =  2 \sum_{k_x > 0, k_y} \Psi_{-\bk} \mathcal{H}_\bk \Psi_{\bk} = 2 \sum_{k_x > 0, k_y} \Psi^\dagger_{\bk} \mathcal{H}_\bk \Psi_{\bk}
\end{align}
The ground state is obtained by filling the negative-energy band,
\begin{align}
    \frac{E_0}{N}  
    =&  \frac{1}{NA} \int_0^{\pi/2} \mathrm{d}k_x \int_{-\frac{\pi}{\sqrt{3}}}^{\frac{\pi}{\sqrt{3}}} \mathrm{d}k_y E^{(-)}_{\bk}
\end{align}
where $E^{(-)}_{\bk} = -4g \sqrt{6 + 2 \cos 2 k_x a + 4 \cos k_x a \cos \sqrt{3} k_y a } $ and  $A = \frac{2}{n} \frac{\pi}{2} \frac{2 \pi}{\sqrt{3}} = 2  \pi^2 \sqrt{3} / N $ is the discretization area in going from the sum to the integral.
Numerically, we find $E_0/N \simeq -1.6 g$. Note that, because the interacting part of the Hamiltonian is positive semidefinite by itself (when including the $\frac{2 N}{3}$ constant), we know that the full ground state energy is bounded by
\begin{equation}
    \frac{E}{N} \geq \frac{E_0}{N} + g^2 = g ( g - 1.6)
\end{equation}
and thus, SUSY is necessarily broken for $g \geq 1.6$.

\section{Generalization of the model}\label{app:general}

Here we consider an extension of the model in Eq.~(\ref{eq:Hamiltonian}), with reversed signs on down-triangle plaquettes,
\begin{align}
    Q = g \sum_j \chi_j + \sum_{\Delta} V_\Delta - \sum_{\nabla} V_\nabla
\end{align}
It this case, the Hamiltonian reads:
\begin{align}
    H =& 2i g \sum_{\langle i,j\rangle} \chi_i \chi_j - 2\sum_{\bowtie} \left( \prod_{j \in \bowtie} \chi_j \right) 
    + \frac{2N}{3} + N g^2
    \label{eq:H_generalized}
\end{align}
where the bilinear terms are now positive in the clockwise direction on up-triangles, and anti-clockwise on down-triangles.
The non-interacting limit has Dirac cones at $\bf{k} = 0$, as shown in Fig.\,\ref{fig:spectrum_generalized}. The infinite-coupling limit ($g=0$) behaves similarly as Eq.~(\ref{eq:Hamiltonian}) -- it only differs by a minus sign in front of the interaction term. In particular, it has the same exponential ground state degeneracy, and can be solved exactly in thin-torus geometries in a similar fashion. The solution still comprises red/green product states, and the blue sector is solved in the same way as presented in the main text. The intermediate-coupling is regime slightly different. The different symmetries lead to different ground-state degeneracy for even systems (16 instead of 8).

\bibliography{main}


\end{document}